\documentclass[nofootinbib,aps,11pt,preprintnumbers]{revtex4}
\pdfoutput=1
\usepackage{graphicx}
\usepackage{cancel}
\usepackage{amssymb}
\usepackage{textcomp}
\usepackage{amsmath}
\usepackage{bm}
\usepackage{times}
\usepackage{epsfig}
\usepackage{color}
\usepackage{graphics}
\usepackage{hyperref}

\begin{document}

\title{\Large {Breaking Local Baryon and Lepton Number at the TeV Scale}}
\author{Pavel Fileviez P\'erez} 
\affiliation{Phenomenology Institute, University of Wisconsin-Madison, 1150 University Ave., Madison, WI 53706, USA }
\author{Mark B. Wise}
\affiliation{California Institute of Technology, Pasadena, CA, 91125 USA}

\date{\today}
\begin{abstract}
Simple models are proposed where the baryon and lepton number are gauged and spontaneously broken near the weak scale.  
The models use a fourth generation that is vector-like with respect to the strong, weak  and electromagnetic interactions to cancel anomalies. 
One does not need large Yukawa couplings  to be consistent with the experimental limits on fourth generation quark masses 
and hence the models are  free of coupling constants with Landau poles near the weak scale.  We discuss the main features of  simple non-supersymmetric and supersymmetric models. 
In these models the light neutrino masses are generated through the seesaw mechanism and proton decay is forbidden even though B and L are broken 
near the weak scale. For some values of the parameters  in these models baryon and/or lepton number violation can be observed at the Large Hadron Collider.  
\end{abstract}
\maketitle

\section{Introduction}
The non-observation of neutrinoless double beta decay, $\tau_{\beta \beta} > 10^{25}$ years~\cite{betabeta}, and proton decay, $\tau_p > 10^{32}$ years~\cite{proton}, indicate that the total lepton number (L)
and the baryon number (B) are very good symmetries in nature. Recently, models have been investigated where baryon number and lepton number are gauged and spontaneously broken at the TeV scale in Refs.~\cite{BL1,BL2,BLSUSY,Ko}.  
Note that by gauging baryon and lepton number we mean that  two new $U(1)$ gauge symmetries are introduced that, for the standard model particles,  correspond to usual baryon and lepton number. In these models  anomalies were cancelled \cite{Foot} by  introducing  a new family (sequential or mirror) with baryon number $\pm 1$ and lepton number $\pm 3$.  These papers build on the earlier work in Refs.~\cite{Foot,Carone}. We view canceling anomalies using a fourth generation as less speculative than other choices  since we  observe this structure for matter. 

The experimental constraints on the masses of fourth generation quarks are strong and so in Refs. ~\cite{BL1,BL2,BLSUSY,Ko} the cutoff of the theory has to be quite low, because of Landau poles in the Yukawa couplings of the fourth generation quarks. In this article, we propose new models where this issue is solved in a simple way. Adding a family that is vector-like (with respect to the $SU(3)\bigotimes SU(2)\bigotimes U(1)$ standard model interactions) and composed  of a chiral family and a mirror family,  we can define an anomaly free theory and generate large masses for the new quarks without assuming large Yukawa couplings. Since baryon and lepton number are vector-like on a complete generation (with a right handed neutrino) we  cancel anomalies  provided the difference in baryon number between the fourth family and mirror family is $-1$ and the difference in lepton number between the fourth family and mirror family is $-3$. The quarks in these families  can have mass terms that do not require weak symmetry breaking provided we introduce a scalar, $S_B$ with baryon number $1$ that gets a vacuum expectation value.

Similarly fourth generation lepton masses that don't require weak symmetry breaking can arise if a scalar $S_L$ with lepton number $3$ gets a vacuum expectation value. We prefer not to take this approach since in this case one 
cannot generate neutrino masses through the seesaw mechanism~\cite{TypeI}. Proton decay is forbidden to all orders if the scalar field that break lepton number has an even lepton number charge. In the models we construct we use fields with $|L|=2$ to break lepton number, generate Majorana neutrino masses through the seesaw mechanism and avoid proton decay. 

In this paper we construct both non-supersymmetric and supersymmetric models where baryon and lepton number are spontaneously broken near the weak scale, proton decay is forbidden and the light neutrinos get mass from the seesaw mechanism. In the non-supersymmetric case we find that after symmetry breaking global baryon number is conserved (by the renormalizable couplings)  while lepton number is broken. In this case observable lepton number violating signals at the LHC from the decays of right-handed neutrinos are possible. In the  supersymmetric scenario the scales for B and L violation are necessarily of order  supersymmetry breaking scale and observable baryon and lepton number violating signals at colliders are possible. Dark matter candidates are briefly discussed in both scenarios. 

This article is organized as follows: In Section II we describe a simple non-supersymmetric model with gauged B and L and without Landau poles in the Yukawa couplings of the new quarks in the theory.
In Section III we discuss the realization of our main idea in a supersymmetric model. Finally, we briefly summarize our main findings in Section IV.

\section{Non-Supersymmetric Scenario} 
%
In order to discuss the spontaneous breaking of gauged B and L we list all Standard Model (SM) fields including their baryonic and leptonic quantum numbers:
$Q_L \sim (3,2,1/6,1/3,0)$, $u_R \sim (3,1,2/3,1/3,0)$, $d_R \sim (3,1,-1/3,1/3,0)$, $l_L \sim (1,2,-1/2,0,1)$, $e_R \sim (1,1,-1,0,1)$ 
and $\nu_R \sim (1,1,0,0,1)$. It is well known that  $U(1)_B$ and $U(1)_L$ are not anomaly free in the SM and to cancel 
anomalies one needs new fermionic degrees of freedom~\cite{Foot}. In Ref.~\cite{Carone} the authors proposed adding a fourth sequential 
family where the new quarks have baryon number $-1$ to cancel all $U(1)_B$ anomalies.  Right handed neutrinos for all the families are required to cancel the lepton number anomalies. Recently, in~\cite{BL1,BL2,BLSUSY} the possibility 
of simultaneously gauging $U(1)_B$ and $U(1)_L$ has been investigated.  The cases of adding a sequential and mirror fourth family were investigated in 
Ref.~\cite{BL1,BL2,BLSUSY} and the main features of the models were discussed.  Experimental constraints on fourth generation quark masses are quite strong and this unfortunately leads to very large fourth family Yukawa couplings that 
have Landau poles located not very far from the weak scale. In this section we  construct 
an anomaly free model based on the gauge group $SU(3)_C \bigotimes SU(2)_L \bigotimes U(1)_Y \bigotimes U(1)_B \bigotimes U(1)_L$ 
free of non-perturbative Yukawa couplings up to the Planck scale. 

\underline{Anomaly Cancellation}: Adding a vector-like family, which is composed of a sequential and mirror family, one can generate large masses for the new fermions in the theory and cancel 
all baryonic and leptonic anomalies. The new fermions are $Q_L^{'}$, $u_R^{'}$, $d_R^{'}$, $l_L^{'}$, $e_R^{'}$ and $\nu_R^{'}$ for the sequential family, 
while one has  $Q_R^{'}$, $u_L^{'}$, $d_L^{'}$, $l_R^{'}$, $e_L^{'}$ and $\nu_L^{'}$ for the mirror family. Anomalies are canceled if 
the following conditions are satisfied,
\begin{eqnarray}
B_{Q_L^{'}}&=& B_{u_R^{'}}=B_{d_R^{'}}, \\
B_{Q_R^{'}}&=& B_{u_L^{'}}=B_{d_L^{'}}, \\
B_{Q_L^{'}}&-& B_{Q_R^{'}}=-1,
\end{eqnarray}
in the quark sector and 
\begin{eqnarray}
L_{l_L^{'}}&=& L_{e_R^{'}}=L_{\nu_R^{'}}, \\
L_{l_R^{'}}&=& L_{e_L^{'}}=L_{\nu_L^{'}}, \\
L_{l_L^{'}}&-& L_{l_R^{'}}=-3,
\end{eqnarray}
in the leptonic sector.

\underline{Quark Sector}: We generate large quark masses without  large Yukawa couplings using the following interactions: 
\begin{eqnarray}
-\Delta {\cal L}_{q' {\rm mass}}&=&h_U^{'} \  \overline{Q^{'}_L} \  \tilde{H} \ u_R^{'}  \ + \  h_U^{''} \  \overline{Q^{'}_R} \  \tilde{H} \ u_L^{'} 
							\ + \  h_D^{'} \  \overline{Q^{'}_L} \  {H} \ d_R^{'}    \ + \  h_D^{''} \  \overline{Q^{'}_R} \  {H} \ d_L^{'}  \nonumber \\
							& + & \lambda_Q  \  \overline{Q^{'}_L} Q_R^{'} S_B \ + \  \lambda_U  \  \overline{u^{'}_R} u_L^{'} S_B \ + \  \lambda_D \  \overline{d^{'}_R}  d_L^{'} S_B 
							\ + \   \rm{h.c.}.
\end{eqnarray}
where the field $S_B$ has baryon number $-1$ and once it gets a vev one breaks $U(1)_B$ generating masses for the new quarks. Neglecting the couplings 
to the SM Higgs,  the new up type quarks form two Dirac fermions with masses $\lambda_Qv_B/\sqrt{2}$ and  $\lambda_Uv_B/\sqrt{2}$ . Similarly the new down type quarks form two Dirac fermions with masses, $\lambda_Qv_B/\sqrt{2}$ and  $\lambda_Dv_B/\sqrt{2}$. Here, 
 $\left< S_B\right>=v_B / \sqrt{2}$ is the vev of the field breaking local baryon number. Also we have adopted a phase convention where the Yukawa couplings, $\lambda_Q$, $\lambda_U$ and $\lambda_D$ are real.

\underline{Leptonic Sector}: The relevant terms to understand the mass generation for the new leptons are given by 
\begin{eqnarray}
-\Delta {\cal L}_{l' {\rm mass}}&=&h_E^{'} \  \overline{l^{'}_L} \  H \ e_R^{'}  \ + \  h_E^{''} \  \overline{l^{'}_R} \  H \ e_L^{'} 
							\ + \  h_\nu^{'} \  \overline{l^{'}_L} \  \tilde{H} \ \nu_R^{'}    \ + \  h_\nu^{''} \  \overline{l^{'}_R} \  \tilde{H} \  \nu_L^{'}  \nonumber \\
							& + & Y_E  \  \overline{l}_L  H e_R \ + \  Y_\nu \  \overline{l}_L  \tilde{H}  \nu_R \ + \  \lambda_{\nu_R} \  \nu_R  \nu_R S_L 
							\ + \   \rm{h.c.},
\end{eqnarray}
where $S_L$ is the a scalar field with lepton number $-2$ and once gets the vev one breaks $U(1)_L$ generating mass for the SM right handed neutrinos.  There is no mixing between the fourth generation and its mirror and between the fourth  (mirror) generation and the usual three generation of leptons. The neutrinos in the fourth generation and its mirror are Dirac particles and presuming that they are lighter than the charged leptons they are stable and are a component of the dark matter.
The masses for the SM light left handed neutrinos are generated through the type I seesaw mechanism in the usual way. The fourth family  (and mirror family) charged leptons get mass from weak symmetry breaking but because the experimental limits on fourth generation charged lepton masses are weaker ($\sim 100 \ {\rm GeV}$) than for quarks strong coupling issues can be avoided. Since $S_L$ has even lepton number charge proton decay is forbidden even if we add non-renormalizable operators to the Lagrangian.

A was discussed in Refs.~\cite{BL1,BL2}, without additional interactions the lightest new quark is stable since there is no mixing between the SM quarks and the new quarks.
Therefore, following the results in Ref.~\cite{BL1} we add a new scalar field $X$ with baryon number, $B_X = 1/3 - B_{Q_R^{'}}$, which has the interactions   
\begin{eqnarray}
-\Delta{\cal L}_{{Y_X}}&=& \lambda_1 \ X \  \overline{Q_L } \ Q_R^{'} \ + \  \lambda_2 \ X \  \overline{u_R} \ u_L^{'} 
\ + \  \lambda_3 \ X \  \overline{d_R } \ d_L^{'} \ + \   \rm{h.c.}.
\label{DM}
\end{eqnarray}
Here, we assume that $X$ does not get a vev.  For a range of parameters it is an acceptable dark matter candidate~\cite{BL2}. It's stability is an automatic consequence of the gauge symmetry and matter content of the model.

The renormalizable interactions in this model conserve a global baryon number where all the quarks have the same charge. However, this symmetry is anomalous and non perturbative instanton effects do not respect it. 
Also non renormalizable interactions that are allowed by the gauge symmetry do violate  baryon number symmetry after spontaneous symmetry breaking. For example,  the dimension 11 operator, $(u_Rd_Rs_R)^2S_B^2 \rightarrow (u_Rd_Rs_R)^2v_B^2/2$,  violates baryon number by two units.

\underline{Symmetry Breaking}: The scalar sector of this model is composed of the fields $H$, $S_L$, $S_B$ and $X$, 
where $S_B$ gets a vev breaking $U(1)_B$. In the leptonic sector since $S_L$ breaks $U(1)_L$  but preserves the discrete ``parity"  $(-1)^{L}$. Hence proton decay is forbidden. One has two extra neutral gauge bosons, $Z_L$ and $Z_B$, in the theory 
associated to the local $U(1)_L$ and $U(1)_B$, respectively. There are some restrictions on the charges of the fields to ensure that it is natural for  $X$  not to have a vev. For recent studies of leptophobic $Z^{'}$s see Ref.~\cite{ZB}.

\underline{Precision Electroweak Constraints}: The model above has a new generation of quarks and leptons and a mirror of this. 
Since the quarks get mass without weak symmetry breaking their masses can be modestly larger than the weak scale and this suppresses 
their contributions to the oblique  precision electroweak parameters, $S$, $T$ and $U$. However, the leptons of the fourth generation and its mirror 
get mass through weak symmetry breaking and we illustrate here that their contribution to these oblique parameters can be acceptably small. 
For our study we use the formulae in Ref.~\cite{Su}. See also Ref.~\cite{EWPO4} for recent studies of the constraints on a fourth 
generation from electroweak precision observables.

\begin{center}
Table I: Oblique precision electroweak variables, $S$, $T$, and $U$ for various values of the fourth generation (denoted by a prime) lepton masses  and the mirror fourth
generation (denoted by double prime) masses.

\vskip0.25in
\begin{tabular}{ccccccc}
\hline
$M_{e'}~~~$ &$M_{\nu'} ~~~ $&$M_{e''}$~~~& $M_{\nu''} $~~~~& $S~~~~$& $T~~~~$ & $U~~~~$  \\ \hline
$120~{\rm GeV}$  & 60~{\rm GeV} & $120~{\rm GeV}$ &60~{\rm GeV}~~~~ & -0.0737 & 0.1267 & 0.0633  \\
$120~{\rm GeV}$  & 60~{\rm GeV} & $120~{\rm GeV}$ &100~{\rm GeV}~~~~ & -0.0022 & 0.0706 & 0.0335  \\
$120~{\rm GeV}$  & 100~{\rm GeV} & $120~{\rm GeV}$ &100~{\rm GeV}~~~~ & 0.0694 &0.0144 & 0.0036  \\
$150~{\rm GeV}$  & 75~{\rm GeV} & $150~{\rm GeV}$ &75~{\rm GeV}~~~~ & -0.0585 & 0.1981 & 0.0516  \\
$150~{\rm GeV}$  & 100~{\rm GeV} & $150~{\rm GeV}$ &100~{\rm GeV}~~~~ & 0.0159 & 0.0893 &0.0167  \\
$150~{\rm GeV}$  & 75~{\rm GeV} & $150~{\rm GeV}$ &100~{\rm GeV}~~~~ & -0.0213 & 0.1437 & 0.0342  \\
$150~{\rm GeV}$  &100~{\rm GeV} & $150~{\rm GeV}$ &125~{\rm GeV}~~~~ & 0.0423 & 0.0559 &0.0099  \\
$120~{\rm GeV}$  & 60~{\rm GeV} & $150~{\rm GeV}$ &75~{\rm GeV}~~~~ & -0.0661 & 0.1624 &0.0575  \\
\hline
\end{tabular}
\end{center}
\vskip0.25in

In Table I we present values for the oblique parameters $S$, $T$ and $U$ for some charged lepton masses less than or equal to $150~{\rm GeV}$. In this region the lepton Yukawa coupling constants for the fourth generation leptons and the mirror leptons remain perturbative up to very high energy scales. The experimental values of the precision electroweak parameters are: $S=0.02\pm 0.11$, $T=0.05\pm 0.12$ and $U= 0.07 \pm 0.12$~\cite{Gfitter}. This fit assumes that the Higgs mass is $120~{\rm GeV} $ and uses the recent experimental values for other standard model parameters.  Note there is a very large positive correlation between the values of $S$ and $T$. So, for example, an up fluctuation in $S$ is likely associated with an up fluctuation in $T$. In Table I we present the values for the S, T and U parameters in some benchmark scenarios. Clearly from the above table we can easily have acceptable precision electroweak fits. For example, in the second and fifth lines $S$ is very small and $T$ is below $0.071$.  

To study further the allowed masses for the new leptons in the theory we show in Figs. 1 and 2 lines of constant $S$ and $T$,  for different values for the lepton masses. In Fig. 1 (left panel) we assume that 
the neutrinos have mass, $M_{\nu'}=M_{\nu''}=100$ GeV while the masses for the charged leptons change between 100 GeV and 200 GeV. In Fig. 1 (right panel) we show the values of $S$, when the charged leptons have a mass equal to 150 GeV 
and the masses for the heavy neutrinos change between 50 GeV and 150 GeV. In both Figures  the values of $S$ are small and are in agreement with the experiment. 
\begin{figure}
\centering
\includegraphics[height=7.9cm,angle=0]{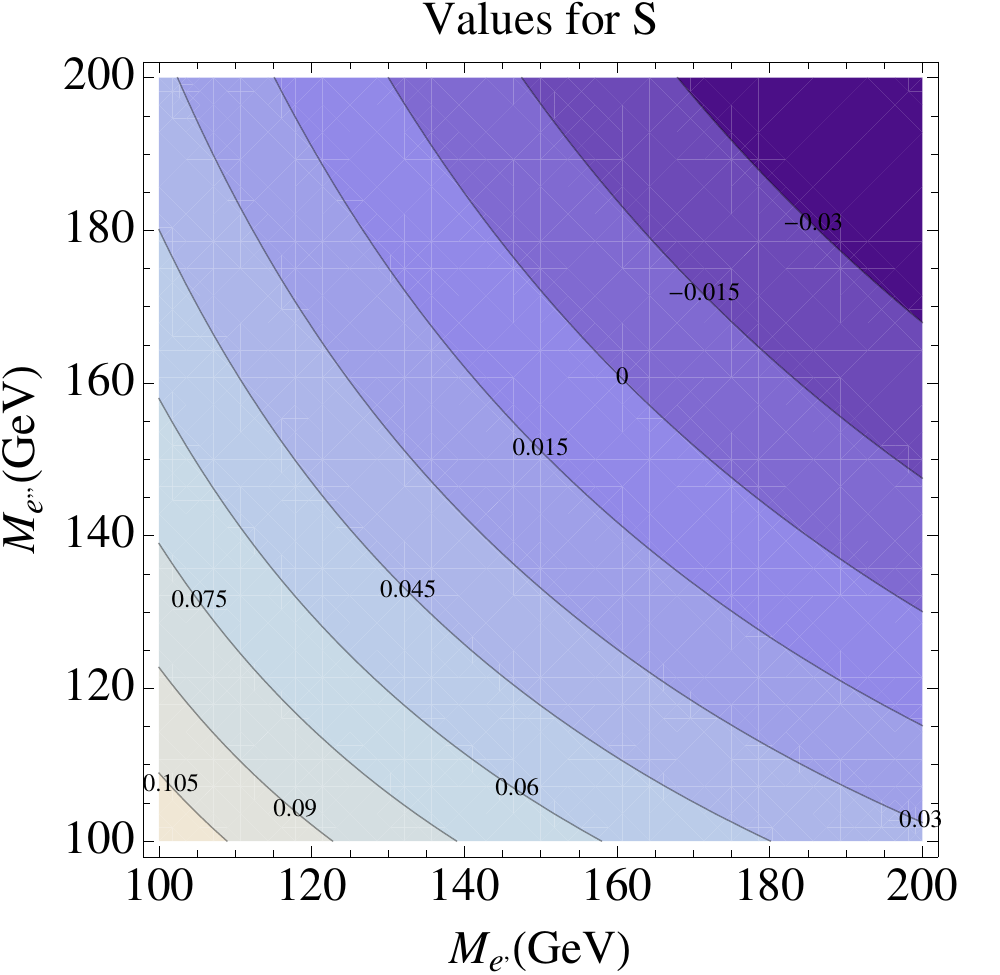}
\includegraphics[height=7.9cm,angle=0]{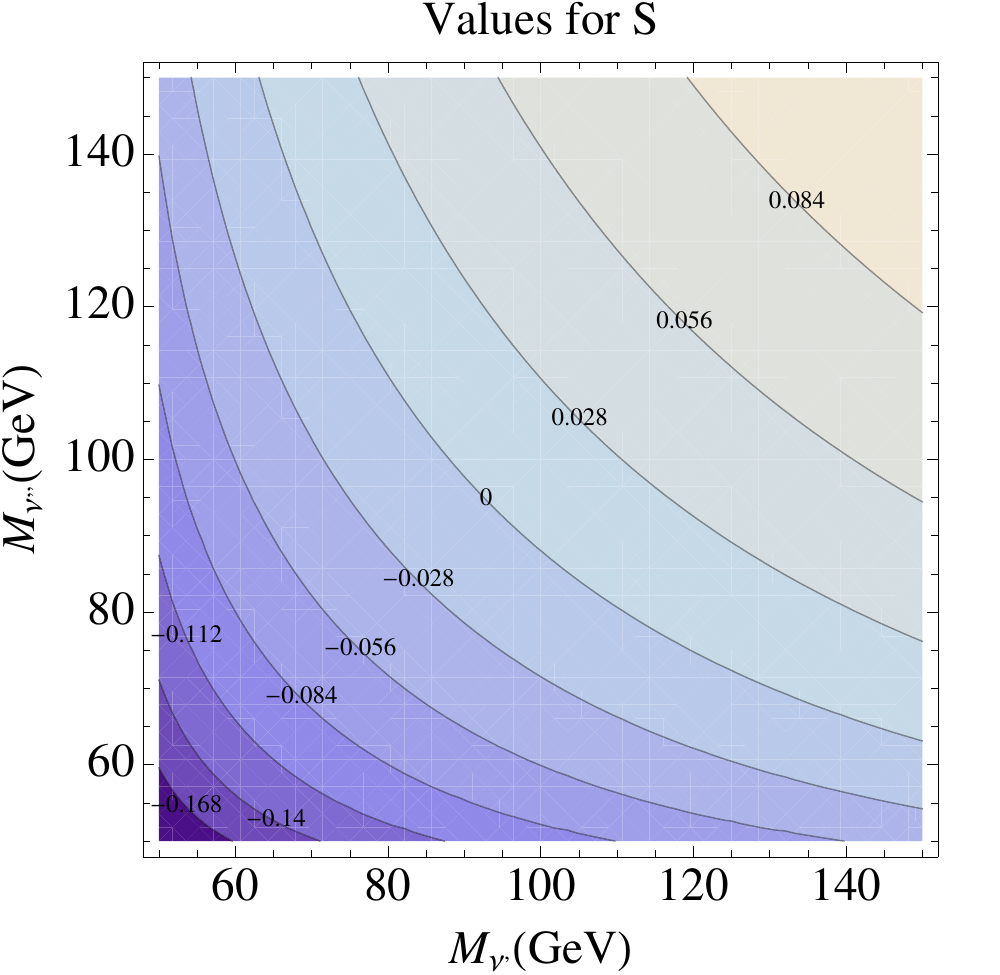}
\caption{Values for the $S$ parameter when the masses of the fourth generation neutrinos is equal to $100$ GeV (left panel) and when the masses of the new charged leptons is 150 GeV (right panel).}
\end{figure}
\begin{figure}
\centering
\includegraphics[height=7.9cm,angle=0]{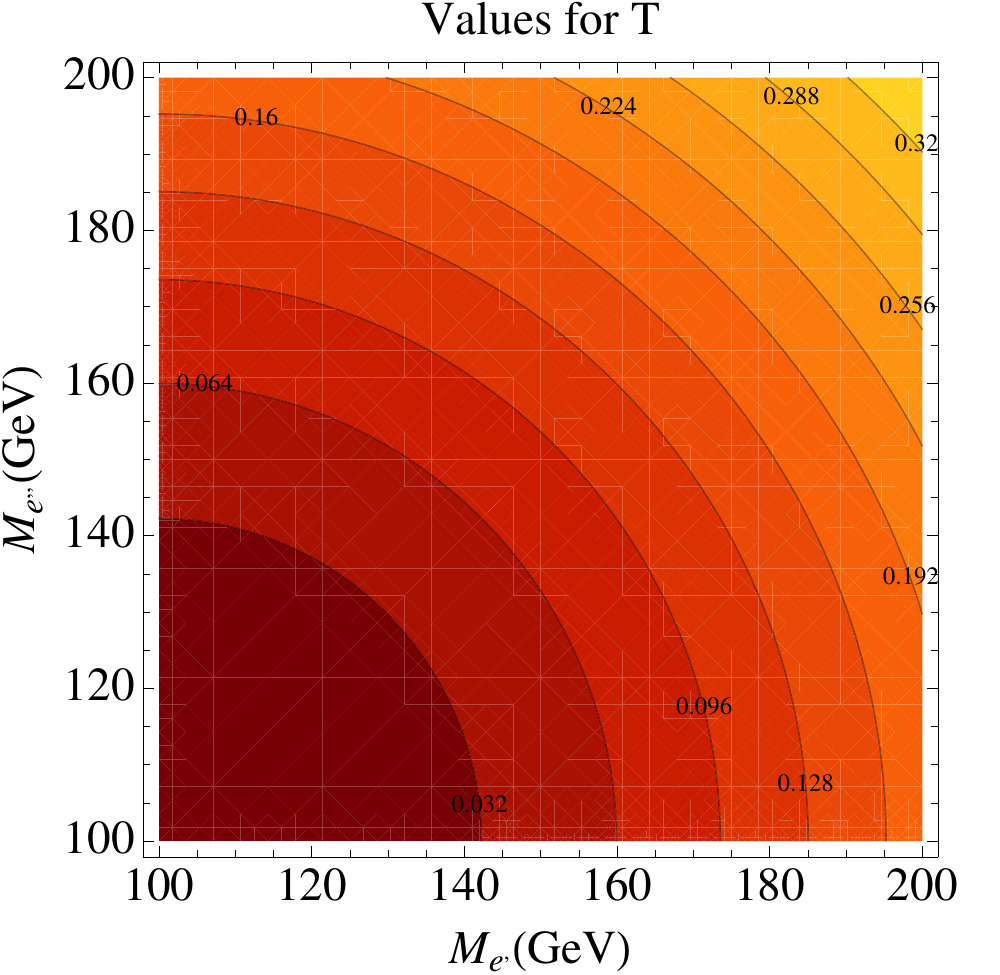}
\includegraphics[height=7.9cm,angle=0]{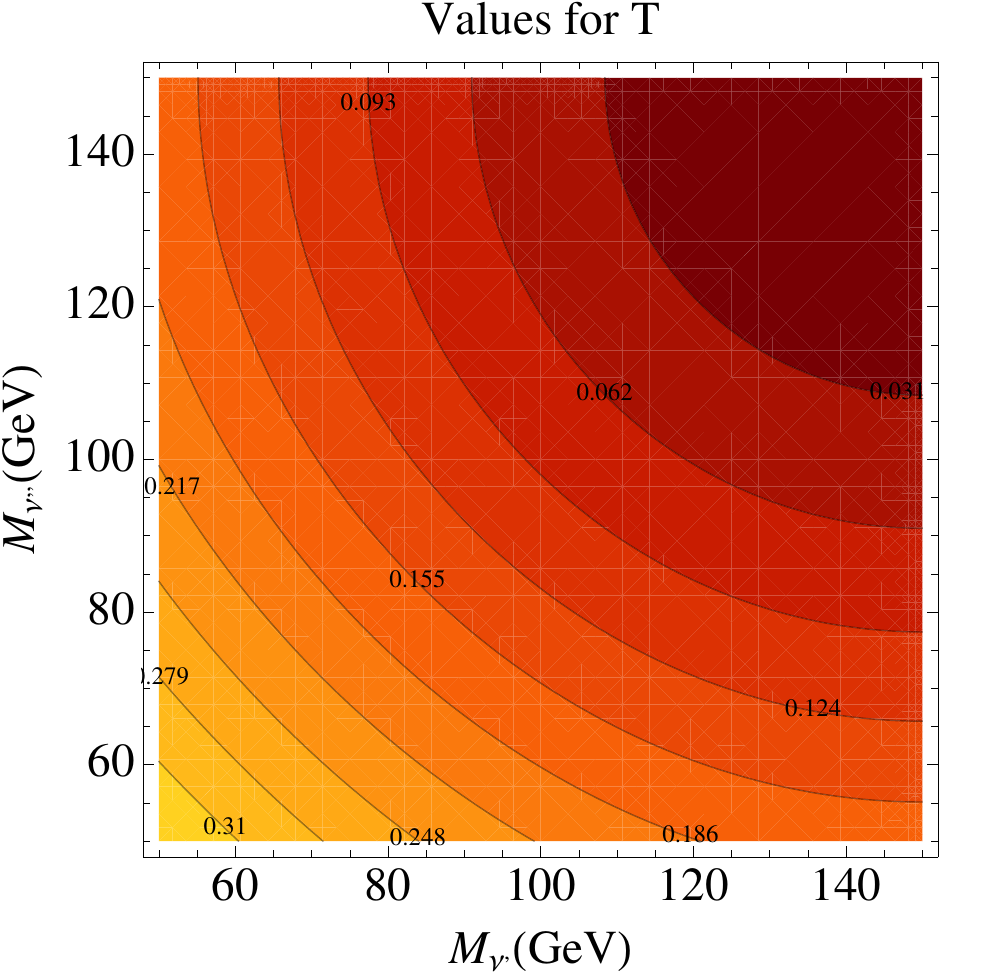}
\caption{Values for the $T$ parameter when the masses of the fourth generation neutrinos are equal to $100$ GeV (left panel) and when the masses of the new charged leptons are 150 GeV (right panel).}
\end{figure}
\begin{figure}
\centering
\includegraphics[height=7.9cm,angle=0]{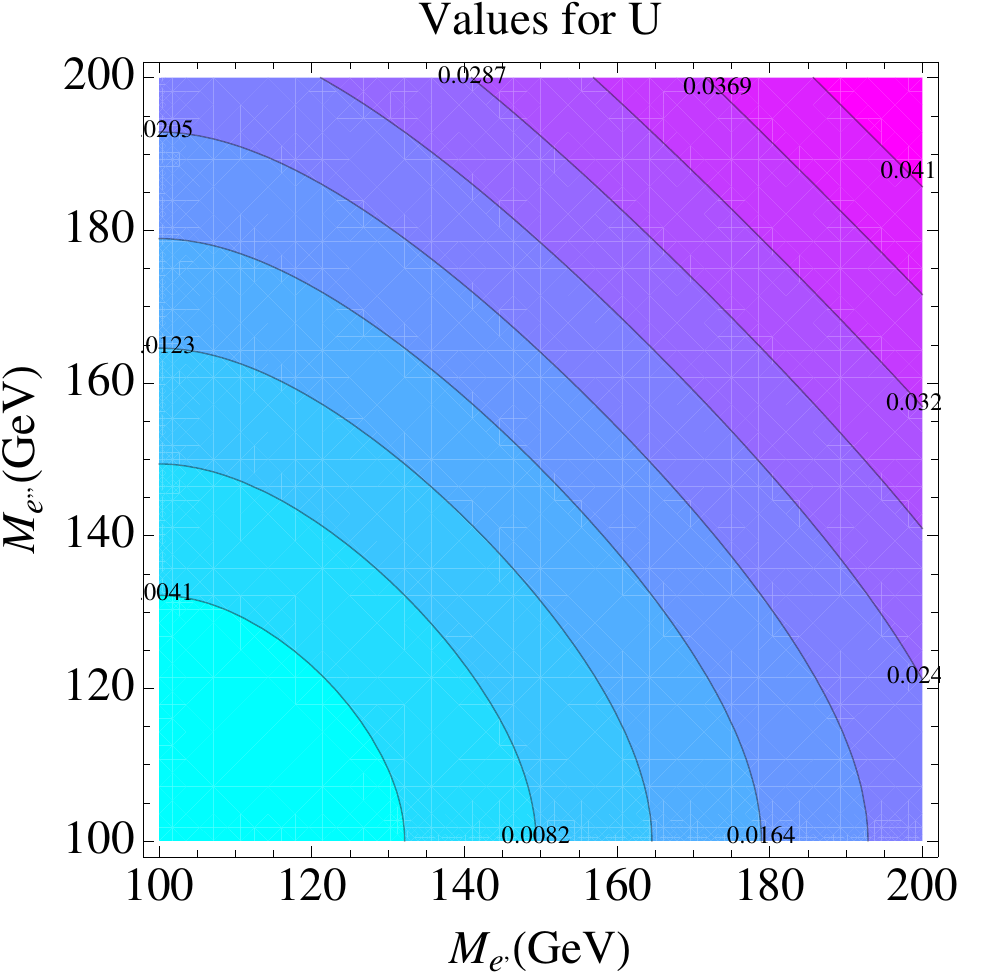}
\includegraphics[height=7.9cm,angle=0]{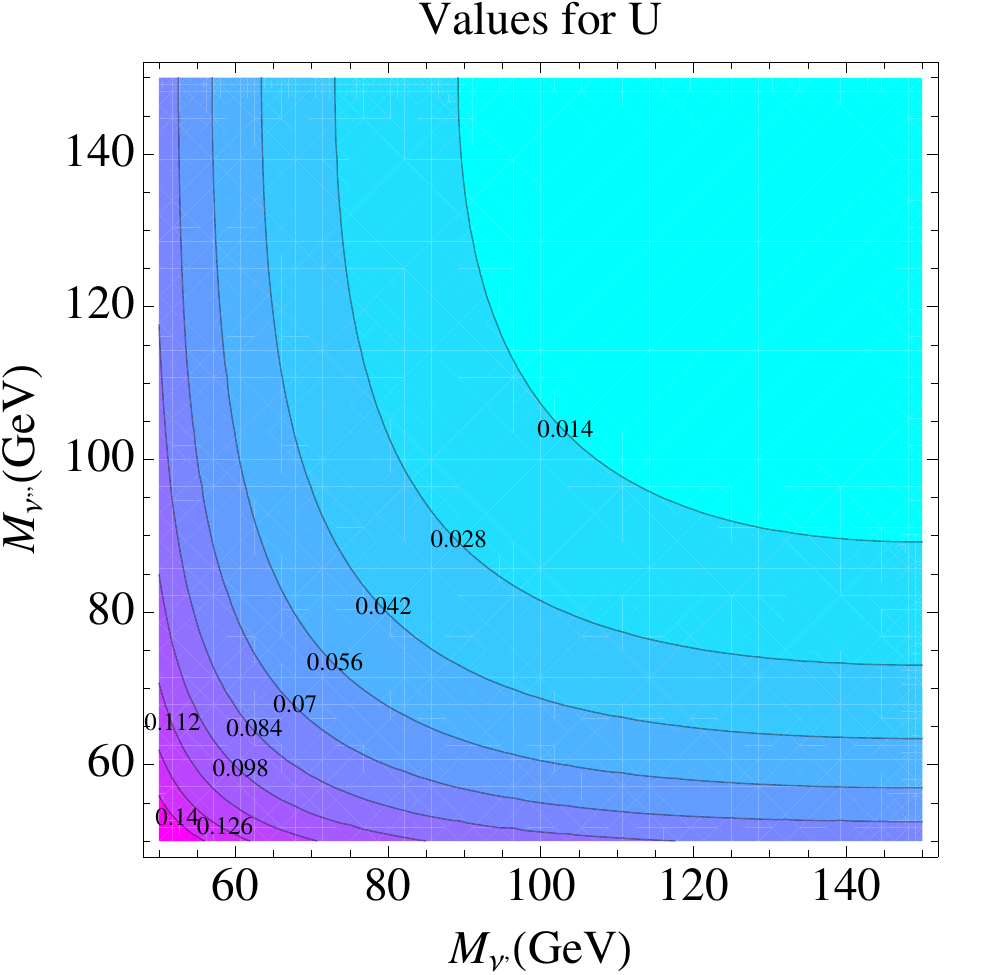}
\caption{Values for the $U$ parameter when the masses of the fourth generation neutrinos are equal to $100$ GeV (left panel) and when the masses of the new charged leptons are 150 GeV (right panel).}
\end{figure}
Next, we study the different values for the $T$ parameter. In Fig. 2 (left panel) we show the values for $T$ assuming  $M_{\nu'}=M_{\nu''}=100$ GeV, and 
one can see that the allowed splitting are, $M_{e'}-M_{\nu'} < 80$ GeV and $M_{e''}-M_{\nu''} < 80$ GeV.  In the second scenario showed in Fig.2 (right panel) we assume that the masses of the charged 
leptons is $150$ GeV. In that case  to have a $T$ parameter below 0.12 requires a mass splitting larger than $70$ GeV between the charged and neutral leptons.

 In Fig. 3 we show  values for the $U$ parameter, which are small over the parameter 
space investigated. These Figures show that there are many regions of parameter space  where 
the new leptons are somewhat light and where the constraints coming from  the measured oblique parameters are satisfied.

\underline{Lepton Number Violation at the LHC}: In this model the first three generation right-handed neutrinos, $\nu_R$, have Majorana mass terms and  for  a range of parameters  lepton number violation is observable at the LHC.
The right handed neutrinos can be produced at the LHC through the $Z^{'}$ gauge bosons which are a linear combination of the $U(1)_B$ and $U(1)_L$ gauge bosons. A large kinetic mixing between 
these two Abelian symmetries is not forbidden by experiment. Then, for example, one has the following signals of lepton number violation: $pp \to Z^{'} \to \nu_R \nu_R \to e^{\pm}_i e^{\pm}_j W^{\mp} W^{\mp} \to e^{\pm}_i e^{\pm}_j 4j$, with same sign dileptons and multijets.
See Ref.~\cite{LNV-LHC} for the study of these channels at the LHC. 

\underline{Flavor Violation}: In this model the new fermions do not mix with the SM fermions at tree level because they have different baryon number in the case of the quarks and lepton number in the case of the leptons. 
Eq.~\ref{DM} induces at  one-loop level mixing between the SM quarks and the new quarks.  Therefore there are new contributions to flavor changing neutral current processes like $K^0-{\bar K}^0$  and $B^0-{\bar B}^0$  mixing. 
These are acceptably small provided that, the couplings $\lambda_ a$ in Eq.~\ref{DM}  are less than $10^{-2}$ ~\cite{BL1}.
   
In this model there is no reason to associate the spontaneous breaking of B and L with the weak scale. Furthermore multiple fine tunings are required to keep all the scalars at that scale. 
Hence it is important to consider a supersymmetric model with spontaneously broken local B and L symmetries where the B and L  symmetry breaking scales must be of order the SUSY breaking scale.   

\section{Supersymmetric Scenario}

In supersymmetric models where B and L are spontaneously broken gauge symmetries  the absence of baryon and lepton number violating
operators of dimension four and five that cause proton decay can be understood. In this section we discuss a simple SUSY model where these symmetries are broken at the TeV scale. 
As usual we define the chiral superfields for the SM fields, $\hat{Q} \sim (3,2,1/6,1/3,0)$, $\hat{u}^c \sim (\bar{3},1,-2/3,-1/3,0)$, $\hat{d}^c \sim (\bar{3},1,1/3,-1/3,0)$, $\hat{L} \sim (1,2,-1/2,0,1)$, 
$\hat{e}^c \sim (1,1,1,0,-1)$, $\hat{\nu}^c \sim (1,1,0,0,-1)$, $\hat{H}_u \sim (1,2,1/2,0,0)$ and $\hat{H}_u \sim (1,2,-1/2,0,0)$, 
and the chiral superfields for the new fermions needed for anomaly 
cancellation, $\hat{Q}_4, \hat{u}_4^c, \hat{d}_4^c, \hat{L}_4, \hat{e}_4^c, \hat{\nu}^c_4$ and $\hat{Q}_4^c, \hat{u}_4, \hat{d}_4, \hat{L}_4^c, \hat{e}_4, \hat{\nu}_4$. 

With this notation  the anomaly cancellation conditions are:
\begin{eqnarray}
B_{Q_4}&=& - B_{u^c_4} = - B_{d^c_4}, \\
B_{Q_4^c} &=& - B_{u_4}= - B_{d_4}, \\ 
B_{Q_4} &+& B_{Q_{4}^c} = -1,
\end{eqnarray} 
in the quark sector and
\begin{eqnarray}
L_{L_4}&=& - L_{e^c_4} = - L_{\nu^c_4}, \\
L_{L_4^c} &=& - L_{e_4}= - L_{\nu_4}, \\ 
L_{L_4} &+& L_{L_{4}^c} = -3,
\end{eqnarray}
in the leptonic sector. 
In this model we have the MSSM superpotential plus the following extra terms relevant for generating masses for the new fermions:
\begin{eqnarray}
{\cal W}_{q_4}&=& Y_U^{'} \  \hat{Q}_4 \hat{H}_u \hat{u}_4^c \ + \  Y_U^{''} \  \hat{Q}_4^c \hat{H}_d \hat{u}_4
			   \ + \ Y_D^{'} \  \hat{Q}_4 \hat{H}_d \hat{d}_4^c \ + \  Y_D^{''}  \  \hat{Q}_4^c \hat{H}_u \hat{d}_4 \nonumber \\
			    &+& \lambda_{Q_4}  \  \hat{Q}_4  \hat{Q}_4^c \hat{S}_B \ + \  \lambda_{u_4}  \  \hat{u}_4^c  \hat{u}_4 \hat{S}_B \ + \  \lambda_{d_4}  \  \hat{d}_4^c  \hat{d}_4 \hat{S}_B.
\end{eqnarray}
The MSSM Higgs superfields are denoted by $\hat{H}_u \sim (1,2,1/2,0,0)$ and $\hat{H}_d \sim (1,2,-1/2,0,0)$. The new Higgs chiral superfields, $\hat{S}_B \sim (1,1,0,1,0)$ and $\hat{\overline{S}}_B \sim (1,1,0,-1,0)$, generate mass for the new leptophobic gauge boson, $Z_B$, and for the new heavy quarks. 

 In the leptonic sector one has the following interactions
\begin{eqnarray}
{\cal W}_{l_4} &=& Y_E^{'}  \  \hat{L}_4 \hat{H}_d \hat{e}^c_4 \ + \  Y_E^{''}  \  \hat{L}_4^c \hat{H}_u \hat{e}_4 \ + \ Y_\nu^{'} \  \hat{L}_4 \hat{H}_u \hat{\nu}^c_4 
                         \ + \ Y_\nu^{''}  \  \hat{L}_4^c \hat{H}_d \hat{\nu}_4  \nonumber \\
                         & + &  Y_{E}  \  \hat{L}  \hat{H}_d  \hat{e}^c  \ + \  Y_{\nu}  \  \hat{L}  \hat{H}_u \hat{\nu}^c 
                         \ + \  \lambda_{\nu^c}  \  \hat{\nu}^c  \hat{\nu}^c  \hat{\overline{S}}_L.
\end{eqnarray}
Here, $\hat{S}_L \sim (1,1,0,0,-2)$ and $\hat{\overline{S}}_L \sim (1,1,0,0,2)$ are chiral superfields that get vacuum expectation values that spontaneously break lepton number and give mass to the quark-phobic gauge boson $Z_L$. Notice that, as in the non-susy case, we have an implementation of the seesaw mechanism for the light neutrino masses.  The new fourth generation neutrinos have Dirac mass terms.

One  difference from the non supersymmetric case is that because the form of the scalar potential is more constrained in SUSY models the symmetry breaking scales for $U(1)_B$ and $U(1)_L$ are necessarily of order  the supersymmetry breaking scale\footnote{We neglect Fayet-Iliopoulos  D terms.}. In order to complete our discussion we list the new superpotential in the Higgs sector:
\begin{eqnarray}
{\cal W}_{S} &=& \mu_{L} \hat{\overline{S}}_L \hat{S}_L \ + \   \mu_{B} \hat{\overline{S}}_B \hat{S}_B.
\end{eqnarray}
\underline{Symmetry Breaking}: As it was discussed in Ref.~\cite{BLSUSY} the symmetries $U(1)_B$ and $U(1)_L$ are broken at the TeV scale since the masses of the new neutral gauge bosons are related  to the SUSY breaking mass scale.
In order to show this we give the dependence of the quarkphobic $Z_L$ gauge boson mass squared on the  parameters in the model:
\begin{equation}
\frac{1}{2} m_{Z_L}^2 = - |\mu_L|^2 \ + \ \left(  \frac{m_{S_L}^2 \tan^2 \beta_L - m_{\bar{S}_L}^2}{ \tan^2 \beta_L - 1 } \right),
\end{equation} 
where $m_{S_L}$ and $m_{\bar{S}_L}$ are soft masses for the Higgses $S_L$ and $\bar{S}_L$, while $\tan \beta_L = \left< S_L\right> / \left< \bar{S}_L\right>$. Note that without the soft  SUSY breaking mass terms  one does not get a positive value of  $m_{Z_L}^2$ which implies that  the lepton number symmetry breaking scale is of order the SUSY breaking scale. 

\underline{Baryon Number Violation}: For any values of the baryonic charges of the new fermions which satisfy the anomaly conditions  the Higgses $\hat{S}_B$ and 
$\hat{\overline{S}}_B$ have charges $1$ and $-1$, respectively. As discussed in Ref.~\cite{BLSUSY}, in this case one can write the following dimension five operator 
which gives rise to baryon number violation
\begin{eqnarray}
\label{bnv}
{\cal W}_B^5 &=& \frac{\tilde{\lambda}^{''}}{\Lambda} \hat{u}^c \hat{d}^c \hat{d}^c \hat{S}_B.
\end{eqnarray}
After  $U(1)_B$ breaking at the TeV scale one has new interactions violating baryon number {\it i.e.},  a squark can decay to two antiquarks.  This renders the lightest neutralino unstable.  
For example, the photino can decay to a quark and a virtual anti squark that then decays into two quarks. Of course if $\lambda''$ is too small and/or $\Lambda$ is too large this  decay will be not 
occur inside an LHC detector and may even give a lifetime for the lightest neutralino that is greater than the age of the universe.

See Refs.~\cite{Goity,Thesis}  for the constraints  on the interactions in Eq.~(\ref{bnv}) coming from dinucleon decays. These give the limit $\tilde{\lambda}^{''}_{uds} v_B / \Lambda < 10^{-8}$.
The limits from $n-\bar{n}$ oscillation experiments are weaker and we do not discuss these here.

 There is no mixing between the SM fermions and the new fermions from renormalizable operators in the model.  At the dimension five level we write the following interactions which  allow the lightest new quark to decay 
\begin{eqnarray}
{\cal W}_{B}^5 &=& \frac{a_1}{\Lambda} \hat{u}^c_4 \hat{d}^c \hat{d}^c \hat{\overline{S}}_B \ + \   \frac{a_2}{\Lambda} \hat{u}^c \hat{d}^c_4 \hat{d}^c \hat{\overline{S}}_B,
\end{eqnarray}
These interactions require $B_{u^c_4}=5/3$.   It is also possible to build models where one does not use non-renormalizable operators to make the lightest fourth generation quark decay, but rather, adds additional fields similar to what was done in the non supersymmetric case.

\underline{Lepton and Baryon Number Violation at the LHC}: The local leptonic symmetry is broken by the vevs of $S_L$ and $\overline{S}_L$ which have $L=-2$ and $L=2$, respectively. 
Therefore, one can only generate lepton number violating interactions  that change lepton number by an even number.  One can look for lepton number violation at the LHC through 
the decays of right handed neutrinos as we discussed in the previous section. 

 If the relevant couplings are not small one can detect baryon number violation in the decays of squarks and gauginos at the LHC.
See Ref.~\cite{Neutralino-BNV} for the study of neutralino decays that violate baryon number.
For example, if we  the gluino is the lightest supersymmetric particle one could have signals with multitops and multibottoms such as $pp \to \tilde{g} \tilde{g} \to t t bb jj$, where j stands for a light jet.

In summary,  we have constructed a supersymmetric  model consistent  with all low energy constraints that has  B and L  spontaneously broken at the SUSY scale. Since one does not need large Yukawa couplings 
for the new quarks and they are in complete representations of $SU(5)$ the  perturbative gauge coupling unification at a high scale is preserved.  It may be possible to generate an acceptable baryon asymmetry through weak scale baryogenesis.
\section{Summary}

We have investigated simple non-supersymmetric and supersymmetric models where local baryon and lepton number symmetries are spontaneously broken near the weak scale.
In order to cancel the anomalies and avoid the Landau poles for the new Yukawa couplings we add a new vector-like family which can be thought of as a sequential 
and mirror family with different baryon and lepton numbers. In these models the light neutrino masses are generated through the seesaw mechanism and proton decay is 
forbidden even if higher dimension operators are included.  However,  baryon number violating nuclear decays are possible. In both models  lepton number violating signals 
at the LHC through the production and decays of right-handed neutrinos are possible. In the case of a small mixing between the two extra Abelian symmetries we have a leptophobic gauge boson 
$Z_B$ and a quarkphobic gauge boson $Z_L$. As we have discussed before, there is no mixing between the Standard Model fermions and the new fermions at tree level, and hence
quite naturally one avoids unacceptably large flavor violation.    

In the non-supersymmetric version of the theory one finds that baryon number is conserved after symmetry breaking and one has a  dark matter candidate with spin zero.
 In the 
supersymmetric version the baryon and lepton number violating scales must be of order the supersymmetry breaking scale. The lightest neutralino is unstable and for a range of parameters  it is possible to observe  baryon number violation through gaugino decays. 
For example, the decays of gluinos may give rise to channels with multi-tops and multi-bottoms which clearly violate baryon number and can be observed at the LHC. 

\textit{Acknowledgments}: {\small We would like to thank N. Arkani-Hamed for pointing out the possibility of using vector like fermions in these scenarios. 
P.F.P. would like to thank the Institute for Advanced Study in Princeton for their hospitality at the beginning of this project.
The work of P.F.P. has been supported in part by the U.S. Department of Energy under grant No. DE-FG02-95ER40896, 
and by the Wisconsin Alumni Research Foundation. The work of M.B.W. was supported in part by the U.S. Department of
Energy under contract No. DE-FG02-92ER40701.}



\begin{thebibliography}{000}

\bibitem{betabeta}
For a review on neutrinoless double beta decay see:
  S.~R.~Elliott and P.~Vogel,
  ``Double beta decay,''
  Ann.\ Rev.\ Nucl.\ Part.\ Sci.\  {\bf 52} (2002) 115
  [arXiv:hep-ph/0202264].

\bibitem{proton}
For a review on proton decay see:
  P.~Nath and P.~Fileviez P\'erez,
  ``Proton stability in grand unified theories, in strings, and in branes,''
  Phys.\ Rept.\  {\bf 441} (2007) 191;
 [arXiv:hep-ph/0601023].


\bibitem{BL1}
  P.~Fileviez P\'erez and M.~B.~Wise,
  ``Baryon and Lepton Number as Local Gauge Symmetries,''
  Phys.\ Rev.\  D {\bf 82} (2010) 011901
  [Erratum-ibid.\  D {\bf 82} (2010) 079901]
  [arXiv:1002.1754 [hep-ph]].


\bibitem{BL2}
  T.~R.~Dulaney, P.~Fileviez Perez, M.~B.~Wise,
  ``Dark Matter, Baryon Asymmetry, and Spontaneous B and L Breaking,''
  Phys.\ Rev.\  {\bf D83 } (2011)  023520.
  [arXiv:1005.0617 [hep-ph]].  
  
\bibitem{BLSUSY}
  P.~Fileviez~Perez and M.~B.~Wise,
  ``Low Energy Supersymmetry with Baryon and Lepton Number Gauged,''
  arXiv:1105.3190 [hep-ph].
  
\bibitem{Ko}
  P.~Ko and Y.~Omura,
  ``Supersymmetric $U(1)_B \times U(1)_L$ model with leptophilic and leptophobic cold
  dark matters,''
  arXiv:1012.4679 [hep-ph].

\bibitem{Foot}
  R.~Foot, G.~C.~Joshi and H.~Lew,
  ``Gauged baryon and lepton numbers,''
  Phys.\ Rev.\  D {\bf 40}, 2487 (1989).

\bibitem{Carone}
  C.~D.~Carone and H.~Murayama,
  ``Realistic models with a light U(1) gauge boson coupled to baryon number,''
  Phys.\ Rev.\  D {\bf 52}, 484 (1995)
  [arXiv:hep-ph/9501220].
  
\bibitem{TypeI}
  P.~Minkowski,
  ``Mu $\to$ E Gamma At A Rate Of One Out Of 1-Billion Muon Decays?,''
  Phys.\ Lett.\ B {\bf 67} (1977) 421;
  T. Yanagida,
in {\it Proceedings of the Workshop on the Unified Theory
   and the Baryon Number in the Universe}, eds. O. Sawada et al.,
p.~95, KEK Report 79-18, Tsukuba (1979);
  M. Gell-Mann, P. Ramond and R. Slansky,
   in {\it Supergravity}, eds. P. van Nieuwenhuizen et al.,
   (North-Holland, 1979), p.~315;
  S.L. Glashow, in {\it Quarks and Leptons}, Carg\`ese, eds. M. L\'evy et al.,
(Plenum, 1980), p. 707;
  R.~N.~Mohapatra and G.~Senjanovi\'c,
  ``Neutrino Mass And Spontaneous Parity Nonconservation,''
  Phys.\ Rev.\ Lett.\  {\bf 44} (1980) 912.
 
  
\bibitem{ZB}
  M.~R.~Buckley, D.~Hooper, J.~Kopp and E.~Neil,
  ``Light Z' Bosons at the Tevatron,''
  arXiv:1103.6035 [hep-ph];
  M.~Buckley, P.~Fileviez Perez, D.~Hooper and E.~Neil,
  ``Dark Forces At The Tevatron,''
  arXiv:1104.3145 [hep-ph];
  F.~Yu,
  ``A Z' Model for the CDF Dijet Anomaly,''
  [arXiv:1104.0243 [hep-ph]];
  K.~Cheung and J.~Song,
  ``Tevatron Wjj Anomaly and the baryonic $Z'$ solution,''
  arXiv:1104.1375 [hep-ph];
  P.~Ko, Y.~Omura and C.~Yu,
  ``Dijet resonance from leptophobic Z' and light baryonic cold dark matter,''
  arXiv:1104.4066 [hep-ph];
  J.~L.~Hewett and T.~G.~Rizzo,
  ``Dissecting the Wjj Anomaly: Diagnostic Tests of a Leptophobic Z',''
  arXiv:1106.0294 [hep-ph].
  
  
  
\bibitem{Su}
  H.~J.~He, N.~Polonsky and S.~f.~Su,
  ``Extra families, Higgs spectrum and oblique corrections,''
  Phys.\ Rev.\  D {\bf 64} (2001) 053004
  [arXiv:hep-ph/0102144].
  
\bibitem{EWPO4}
  J.~Erler and P.~Langacker,
  ``Precision Constraints on Extra Fermion Generations,''
  Phys.\ Rev.\ Lett.\  {\bf 105} (2010) 031801
  [arXiv:1003.3211 [hep-ph]];
  O.~Eberhardt, A.~Lenz and J.~Rohrwild,
  ``Less space for a new family of fermions,''
  Phys.\ Rev.\  D {\bf 82} (2010) 095006
  [arXiv:1005.3505 [hep-ph]];
  M.~Bobrowski, A.~Lenz, J.~Riedl and J.~Rohrwild,
  ``How much space is left for a new family of fermions?,''
  Phys.\ Rev.\  D {\bf 79} (2009) 113006
  [arXiv:0902.4883 [hep-ph]];
  S.~Dawson and P.~Jaiswal,
  ``Four Generations, Higgs Physics, and the MSSM,''
  Phys.\ Rev.\  D {\bf 82} (2010) 073017
  [arXiv:1009.1099 [hep-ph]];
  G.~D.~Kribs, T.~Plehn, M.~Spannowsky and T.~M.~P.~Tait,
  ``Four generations and Higgs physics,''
  Phys.\ Rev.\  D {\bf 76} (2007) 075016
  [arXiv:0706.3718 [hep-ph]].
  
  
  \bibitem{Gfitter}
  http://gfitter.desy.de/GOblique/
  
\bibitem{LNV-LHC}
  K.~Huitu, S.~Khalil, H.~Okada and S.~K.~Rai,
  ``Signatures for right-handed neutrinos at the Large Hadron Collider,''
  Phys.\ Rev.\ Lett.\  {\bf 101} (2008) 181802
  [arXiv:0803.2799 [hep-ph]];
  J.~A.~Aguilar-Saavedra,
  ``Heavy lepton pair production at LHC: Model discrimination with multi-lepton
  signals,''
  Nucl.\ Phys.\  B {\bf 828} (2010) 289
  [arXiv:0905.2221 [hep-ph]];
  P.~Fileviez Perez, T.~Han and T.~Li,
  ``Testability of Type I Seesaw at the CERN LHC: Revealing the Existence of
  the B-L Symmetry,''
  Phys.\ Rev.\  D {\bf 80} (2009) 073015
  [arXiv:0907.4186 [hep-ph]].
  
  
\bibitem{Goity}
  J.~L.~Goity and M.~Sher,
  ``Bounds on delta B = 1 couplings in the supersymmetric standard model,''
  Phys.\ Lett.\  B {\bf 346} (1995) 69
  [Erratum-ibid.\  B {\bf 385} (1996) 500]
  [arXiv:hep-ph/9412208];

\bibitem{Thesis}
Michael D. Litos, ``A search for dinucleon decay into kaons using the SK water cherenkov detector",
Ph.D. Thesis, Boston University, 2010.

  
  \bibitem{Neutralino-BNV}
  J.~M.~Butterworth, J.~R.~Ellis, A.~R.~Raklev, G.~P.~Salam,
  ``Discovering baryon-number violating neutralino decays at the LHC,''
  Phys.\ Rev.\ Lett.\  {\bf 103}, 241803 (2009).
  [arXiv:0906.0728 [hep-ph]].



\end{thebibliography}
\end{document}